\documentstyle[preprint,aps,floats]{revtex}
\input psfig.tex 
\tightenlines

\def\be{\begin{equation}}
\def\ee{\end{equation}}
\def\ba{\begin{eqnarray}}
\def\ea{\end{eqnarray}}
\def\bibi{\bibitem}

\title{{ \bf Why Does Inflation Start at the Top of the Hill?}}

\author{{\bf S.W. Hawking}\thanks{email: S.W.Hawking@damtp.cam.ac.uk},
{\bf Thomas Hertog}\thanks{email: T.Hertog@damtp.cam.ac.uk}, \\ 
DAMTP, Centre for Mathematical Sciences, University of
Cambridge \\ Wilberforce Road, Cambridge CB3 0WA, United Kingdom.}

\date{\today}
\begin{document}
\maketitle

\begin{abstract}

We show why the universe started in an unstable de Sitter state.
The quantum origin of our universe implies one must take a `top down'
approach to the problem of initial conditions in cosmology, in which
the histories that contribute to the path integral, depend on the
observable being measured. Using the no boundary
proposal to specify the class of histories, we study the quantum
cosmological origin of an inflationary
universe in theories like trace anomaly driven inflation in which the
effective potential has a local maximum.  We find that
an expanding universe is most likely to emerge in an unstable
de Sitter state, by semiclassical tunneling via a Hawking-Moss instanton.
Since the top down view is forced upon us by the quantum nature of the
universe, we argue that the approach developed here should still apply
when the framework of quantum cosmology will be based on M-Theory.

\end{abstract}

\section{Introduction}

Structure and complexity have developed in our universe, because 
it is out of equilibrium. 
This feature shows up in all known cosmological scenarios for the
early universe, which rely on gravitational instability to generate local
inhomogeneities from an almost homogeneous and isotropic state for the
universe.  Inflation seems the best explanation for this homogeneous and
isotropic state because whatever drives the inflation will remove the local
instability and iron out irregularities. However the inflationary expansion
has to be globally unstable because otherwise it would continue forever and
galaxies would never form.

The instability can be described as the evolution of an order parameter
$\phi $ which can be treated as a scalar field with effective potential
$V(\phi )$. If $ V' /V$ is small, $\phi $ will roll slowly down the
potential and the universe will inflate by a large factor. However, this
raises the question: Why did the universe start with a high value of the
potential? Why didn't  $\phi $ start at the global minimum of $V$?

There have been various attempts to explain why $\phi $ started high on the
potential hill. In the old \cite{Guth82} and new \cite{Linde82,Albrecht82}
inflationary scenarios the universe was
supposed to start with infinite temperature at a singularity. As the
universe expanded and cooled, thermal corrections would make the effective
potential time dependent. So even if $\phi $ started in the minimum of $V$,
it could still end up in a metastable false vacuum state (in old inflation)
or at a local maximum of $V$ (in new inflation). The scalar field was then
supposed to tunnel through the potential barrier or just fall off the top
of the hill and slowly roll down. However both scenarios tended to predict
a more inhomogeneous universe than we observe. They were also
unsatisfactory because they assumed an initial singularity and a fairly
homogeneous and isotropic pre-inflation hot big bang phase. Why not just
assume the singularity produced the standard hot big bang, since we don't
have a measure on the space of singular initial conditions for the
universe?

In the chaotic inflation scenario \cite{Linde83}, 
quantum fluctuations of $\phi $ are
supposed to drive the volume weighted average $\phi $ up the potential 
hill, leading to everlasting eternal inflation. However this effect is
dependent on using the synchronous gauge: in other gauges the volume
weighted average of the potential can go down. Looking from a 4 rather than
3+1 dimensional perspective, it is clear that the quantum fluctuations of a
single scalar field are insufficient to drive de Sitter like eternal
inflation, if the de Sitter space is larger than the Planck length. Eternal
inflation may be possible at the Planck scale, but all our methods would
break down in this situation so it would mean that we could not analyze the
origin of the universe.

The aim of this paper however is to show that the universe can come into
being and start inflating without the need for an initial hot big bang
phase or Planck curvature. It is required that the potential $V$ has a
local maximum which is below the Planck density and sufficiently flat on
top, $V''/V >-4/3$. This last condition means only the homogeneous mode of the
scalar field is tachyonic: the higher modes all have positive eigenvalues.
It also means there isn't a Coleman-De Luccia solution \cite{Coleman80}
describing quantum tunneling from a false vacuum on one side of the maximum 
to the true vacuum on the other side. Instead there is only a homogeneous 
Hawking-Moss instanton \cite{Hawking82a}
that sits on the top of the hill, at the local maximum of $ V $.

It has long been a problem to understand how the universe could decay from a
false vacuum in this situation. The Hawking-Moss instanton does not
interpolate between the false and true vacua, because it is constant in
space and time. Instead, what must happen is that the original universe can
continue in the false vacuum state but that new completely disconnected
universes can form at the top of the hill via Hawking-Moss instantons. For
someone in one of these new universes, the universe in the false vacuum is
irrelevant and can be ignored.

The top of the hill might seem the least likely place for the universe to
start. However we shall show it is the most likely place for an
inflationary universe to begin, if $V''/V >-4/3$. 
The reason is that although being at
the top of the hill costs potential action, the saving of gradient action
from having a constant scalar field is greater.  Thus inflation will start
at the top of the hill. In particular, this justifies Starobinsky's scenario
of trace anomaly inflation, in which the universe starts in an unstable de 
Sitter state supported by the conformal anomaly of a large number of 
conformally coupled matter fields \cite{Starobinsky80}.

The usual approach to the problem of initial conditions for inflation, is to
assume some initial configuration for the universe, and evolve it forward 
in time. This could be described as the bottom up approach to cosmology.
It is an essentially classical picture, because it assumes there is a single
well defined metric for the universe.  By contrast, here we adopt a quantum
approach, based on the no boundary proposal \cite{Hartle83},
which states that the amplitude for an observable like the 3-metric on a
spacelike hypersurface $\Sigma$, is given by a path integral over all
metrics
whose only boundary is $\Sigma$. The quantum origin of our universe
and the no boundary proposal naturally lead to a
top down view of the universe, in which the histories that contribute to
the path integral, depend on the observable being measured.

We study the quantum cosmological origin of an expanding universe in
theories like trace anomaly inflation,
by investigating the semiclassical predictions of the no boundary
proposal for the wave function of interest.
One may argue that a clearer picture of the pre-inflationary conditions
can only emerge from a deeper understanding of quantum gravity at the Planck
scale. However, the amplitude of the cosmic microwave temperature
anisotropies
indicates that the universe may always have been much larger than the Planck
scale. This suggests it might be possible to describe the origin of our
universe within the semiclassical regime of quantum cosmology.
Correspondingly, the effective potential must have a local maximum 
well below the Planck density, which is the case in the trace anomaly model.

The paper is organised as follows. In section 2 we review trace anomaly 
driven inflation, since this provides an important theoretical motivation 
for inflation. We study the quantum cosmology of the trace anomaly model and
discuss the role of a special class of instanton saddle-points
of the no boundary path integral, which can be analytically
continued to Lorentzian universes.
In section 3 we consider perturbations in anomaly-induced inflation and show
that the instability of the inflationary phase can be
described by a scalar field with an effective potential with a local maximum.
We also discuss homogeneous fluctuations about the instanton backgrounds and
touch briefly on the effect of quantum matter on the
spectrum of microwave fluctuations predicted by anomaly-induced inflation.
In section 4, we consider a general model of inflation with 
an effective potential that has a local maximum. We show that according
to the no boundary proposal, provided the instability is sufficiently weak, 
an expanding universe is most likely to start at the top of the hill, 
in a de Sitter state. Finally, in section 5 we present our conclusions.

\section{Trace Anomaly Driven Inflation}

\subsection{Large $N$ Cosmology}

It has been argued that the theoretical foundations for inflation are weak, 
since it has proven difficult to realise inflation in classical M-theory. 
A large 
class of supergravity theories admit no warped de Sitter compactifications 
on a compact, static internal space \cite{Gibbons85,Maldacena01a} and although
some gauged $N=8$ and $N=4$ supergravities in $D=4$ do permit de Sitter vacua 
\cite{Gates83,Hull84}, these vacua are too unstable for a significant period 
of inflation to 
occur. However, an appealing way to evade the no go theorems is to include 
higher derivative quantum corrections to the classical supergravity equations, 
such as the trace anomaly. 

Since we observe a large number of matter fields in the universe, 
it is natural to consider the large $N$ approximation \cite{Tomboulis77}.
In the large $N$ approximation, one performs the path integral over the matter 
fields in a given background to obtain an effective action that is a
functional of the background metric,
\be
 \exp(-W[{\bf g}]) = \int d[\phi] \exp (-S[\phi; {\bf g}]).
\ee
In the leading-order $1/N$ approximation, one can neglect graviton loops and 
look for a stationary point of the effective action for the matter fields 
combined with the gravitational action. This is equivalent to solving the 
Einstein equations with the source being the expectation value of the matter 
energy-momentum tensor derived from $W$,
\be
 R_{\mu\nu} - \frac{1}{2} R g_{\mu\nu} = 8\pi G \langle T_{\mu\nu} \rangle.
\ee
The expectation value of the energy-momentum tensor is generally
non-local and depends on the quantum state. However, during inflation, 
particle masses are small compared with the 
spacetime curvature, $R >>m^2$, and in asymptotically free gauge theories, 
interactions become negligible in the same limit.
Therefore, at the high curvatures during inflation, the energy-momentum tensor 
of a large class of grand unified theories is to a good approximation given 
by the expectation
value $ \langle T_{\mu\nu} \rangle$ of a large number of free, massless, 
conformally invariant fields\footnote{For simplicity, it is
assumed that scalar fields become conformally coupled at high energies, but the
contribution of the interaction terms to $\langle T_{\mu\nu} \rangle$
is small at high curvature, as long as the couplings don't become very 
large \cite{Parker84}.}. The entire one-loop 
contribution to the trace of the energy-momentum tensor then comes from the 
conformal anomaly \cite{Capper74}, which is given for a general CFT 
by the following equation, 
\be
\label{eqn:anomform}
g^{\mu\nu} \langle T_{\mu\nu} \rangle = c F - a G + \alpha' \nabla^2 R,
\ee
where $F$ is the square of the Weyl tensor,
$G$ is proportional to the Euler density
and the constants $a,c$ and $\alpha'$ are given in terms of the field
content of the CFT by
\be
 a = \frac{1}{360(4\pi)^2} \left( N_S + 11 N_F + 62 N_V \right),
\ee
\be
 c = \frac{1}{120(4\pi)^2} \left( N_S + 6N_F + 12 N_V \right), 
\ee
\be
\alpha' = \frac{1}{180(4\pi)^2} \left (N_S + 6 N_F - 18 N_V \right),
\ee
with $N_S$ the number of real scalar fields, $N_F$ the number of
Dirac fermions and $N_V$ the number of vector fields\footnote{We have quoted 
the value for $\alpha'$ predicted by AdS/CFT, which agrees with 
point-splitting or zeta function regularisation \cite{Birrell82}.}.

The trace anomaly is entirely geometrical in origin and therefore independent
of the quantum state. In a maximally symmetric spacetime, 
the symmetry of the vacuum implies
that the expectation value of the energy-momentum tensor is proportional 
to the metric,
\be
\langle 0| T_{\mu\nu} |0 \rangle = \frac{1}{4} g_{\mu\nu} g^{\rho\sigma}
\langle 0| T_{\rho\sigma} |0\rangle.
\ee
Thus the trace anomaly acts just like a cosmological constant for
these spacetimes, and a positive trace anomaly permits a de Sitter
solution to the Einstein equations. 

The radius of the de Sitter solution is determined by the number of 
fields, $N^2$, in the CFT and is of order $ \sim Nl_{pl}$. 
Therefore the one-loop 
contributions to the energy-momentum tensor are $\sim 1/N^2$, which means they
are of the same order as the classical terms in the Einstein equations. On the 
other hand, the corrections due to graviton loops are $\sim 1/N^3$, so for 
large $N$ quantum gravitational fluctuations are suppressed, confirming the 
consistency of the large $N$ approximation. 

For $\alpha'=0$ in \ref{eqn:anomform}, the only $O(3,1)$ invariant
solutions are de Sitter space and flat space, which are the initial and final 
stages of the simplest inflationary universe. In order for a solution
to exist that interpolates between these two stages, one must have
$\alpha' <0$ in \ref{eqn:anomform}, as Starobinsky discovered 
\cite{Starobinsky80}. Starobinsky showed that if $\alpha' <0$, the de
Sitter solution is unstable, and decays into 
a matter dominated Friedman-Lemaitre Robertson-Walker universe, on a timescale 
determined by $\alpha'$. The purpose of Starobinsky's work was to 
demonstrate that quantum effects of matter fields might resolve the Big Bang 
singularity. From a modern perspective, it is more interesting that the 
conformal anomaly might have been the source of a finite, but
significant period of inflation in the early universe. Rapid
oscillations in the expansion rate at 
the end of inflation, would result in particle production and (p)reheating.

Starobinsky showed that the de Sitter solution is unstable both to the future
and to the past, so it was not clear how the universe could have entered the
de Sitter phase. This is the problem of initial conditions for trace anomaly 
driven inflation, which should be addressed within the framework of
quantum cosmology, by 
combining inflation with a theory for the wave function $\Psi$ of the quantum
universe. Hartle and Hawking suggested that the amplitude for 
the quantum state of the universe described by 3-metric ${\bf h}$ and matter 
fields $\phi ({\bf x})$ on a 3-surface $\Sigma$, should be given by
\be\label{nbp}
\Psi [\Sigma,{\bf h},\phi_{\Sigma}] =
N \sum_{M} \int {\cal D} [{\bf g}] {\cal D}[\phi ({\bf x})]
e^{-S_{E}({\bf g},\phi)},
\ee
where the Euclidean path integral is taken over all compact four
geometries 
bounded 
only by a 3-surface $\Sigma$, with induced metric ${\bf h}$ and matter fields
$\phi_{\Sigma}$. $M$ denotes a diffeomorphism class of 4-manifolds and
$N$ is a normalisation factor. 
The motivation to restrict the class of manifolds and metrics 
to geometries with only
a single boundary is that in cosmology, in contrast with scattering 
calculations, one is interested in measurements in a finite region in
the interior of spacetime. The `no boundary' proposal gives a definite
ansatz for the wave function $\Psi [\Sigma, {\bf h}, \phi_{\Sigma}]$ of the
universe and in principle removes the initial singularity in the hot Big Bang 
model. At least within the semiclassical regime, this yields a well-defined 
probability measure on the space of initial conditions for cosmology. 

One can appeal to quantum cosmology to explain how the de Sitter phase
emerges in trace anomaly inflation, since the no boundary proposal 
can describe the creation of an inflationary universe from nothing. At 
the semiclassical level, this process is mediated by a compact instanton 
saddle-point of the Euclidean path integral, which extrapolates to a
real Lorentzian universe at late times. 
To find the relative probability of different geometries in the no
boundary path integral, one must compute
their Euclidean action. In the next section, we consider a model of 
anomaly-induced inflation consisting of gravity coupled to ${\cal N}=4$, 
$U(N)$ super Yang-Mills theory, for which the AdS/CFT correspondence
\cite{Maldacena98} provides an
attractive way to calculate the effective matter action on backgrounds without
symmetry. The fact that we are using ${\cal N}=4$, $U(N)$ super Yang-Mills 
theory is probably not significant since, 
as we shall describe, it is the large number of fields that matters in our 
discussion and not the Yang-Mills coupling. 
Therefore, we expect our results to be valid for any matter theory that is 
approximately massless during the de Sitter phase.

\subsection{Effective Matter Action}\label{effact}

We consider, in Euclidean signature, Einstein
gravity coupled to a ${\cal N}=4$, $U(N)$ super Yang-Mills theory with
large $N$,
\be\label{tot}
 S = -\frac{1}{2 \kappa} \int d^4 x \sqrt{g} R - \frac{1}{\kappa}
\int d^3 x \sqrt{h} K + W,
\ee
where $W$ denotes the Yang-Mills effective action.
The field content of the Yang-Mills theory is
$N_S = 6N^2,\ N_F = 2N^2$ and $N_V = N^2$, yielding an anomalous trace
\be
g^{\mu\nu} \langle T_{\mu\nu} \rangle = \frac{N^2}{64\pi^2} (F-G).
\ee
The one-loop result for the conformal anomaly is exact, since it is protected 
by supersymmetry. Therefore, inflation supported by the trace 
anomaly of ${\cal N}=4$, $U(N)$ super Yang-Mills would never end. 
The presence of non-conformally invariant fields in
realistic matter theories, however, necessarilly alters the value of
$\alpha'$ in the anomaly \ref{eqn:anomform}. 
Since the coefficient of the $\nabla^2 R$ term plays 
such an important role in trace anomaly driven inflation, we ought to include
this correction. As a first approximation, one can account for
the non-conformally invariant fields by adding a
local counterterm to the action, 
\be
S_{ct} =\frac{\alpha N^2}{192 \pi^2} \int d^4 x 
\sqrt{g} R^2. 
\ee
This leads to an extra contribution to the
conformal anomaly, which becomes
\be
g^{\mu\nu} \langle T_{\mu\nu} \rangle = \frac{N^2}{64\pi^2} (F-G)
+\frac{\alpha N^2}{16 \pi^2} \nabla^2 R.
\ee
For $\alpha <0$, the expansion now changes from exponential 
to the typical power law $ \sim t^{2/3}$ of a matter dominated universe, 
on a time scale $\sim 12 \vert \alpha \vert \log N$. 
One can construct more sophisticated models of 
anomaly driven inflation, by taking in account corrections from particle 
masses and interactions in a more precise way. One could, for instance, 
consider soft supersymmetry breaking during inflation. 
The coefficient $\alpha'$ could then vary in time, 
because the decoupling of massive sparticles at low energy 
\cite{Shapiro01} alters the number of degrees of freedom that contribute
to the quantum effective action. 
For our purposes, however, it is sufficient to consider the
theory above.

In no boundary cosmology, one is interested in solutions that 
describe a Lorentzian inflationary universe that emerges from a compact
instanton solution of the Euclidean field equations. These geometries provide
saddle-points of the Euclidean path integral \ref{nbp} for the wave function
of interest. 
Because our universe is Lorentzian at late times, it has been suggested that 
the relevant instanton saddle-points of the no boundary path integral
are so-called `real tunneling' geometries \cite{Gibbons90,Halliwell90a}. 
Cosmological real tunneling solutions are compact Riemannian geometries joined 
to an $O(3,1)$ invariant Lorentzian solution of Einstein's equations,
across a hypersurface of vanishing extrinsic curvature $K_{\mu\nu}$.
Such instanton solutions can then be used as background in a 
perturbative evaluation of the no boundary path integral, to find correlators
of metric perturbations during inflation, 
which in turn determine the cosmic microwave anisotropies.

We now compute the effective matter action $W$ on such perturbed instanton metrics.
After eliminating the gauge freedom, the perturbed metric on the spaces of 
interest can be written as
\be \label{Bmet}
ds^2 = B^2 (\chi) \gamma_{\mu \nu} dx^{\mu}dx^{\nu} =
B^2 (\chi)( (1 +\psi) \hat \gamma_{\mu \nu} +\theta_{\mu \nu})
dx^{\mu}dx^{\nu}, 
\ee
where $\hat \gamma_{\mu \nu}$ is the metric on the unit $S^4$ and 
$\theta_{\mu \nu}$ is transverse and traceless with respect to the four sphere.

In order to evaluate the no boundary path integral, we must first compute the 
quantum effective action $W[B,{\bf h}]$ on the background \ref{Bmet}. 
The effective action of the matter fields is computed as an 
expansion around the homogeneous background with metric
$g_{\mu \nu} = B^2 (\chi)\hat 
\gamma_{\mu \nu}$. To second order in the metric perturbation, 
$W[B,{\bf h}]$ is determined by the one and two-point function of the
energy-momentum tensor on the unperturbed $O(4)$ invariant background.
The one-point function is given by the conformal anomaly. 
Since the FLRW background is conformal to the round four sphere,
the two-point function can be calculated by a conformal transformation 
from $S^4$. On $S^4$, the two-point function is determined entirely
by symmetry and the trace anomaly \cite{Hawking01a}. Therefore, since
the energy-momentum tensor transforms anomalously, the two-point function
on \ref{Bmet} should be fully determined by the two-point function on $S^4$, 
the trace anomaly and the scale factor $B(\chi)$. For the  matter
theory we have in mind, all these quantities are independent of the 
coupling, so it follows that the effective action $W[B,{\bf h}]$ is
independent of the coupling, to second order in the metric perturbation.
 
In \cite{Riegert84}, it was found how the effective action that generates a 
conformal anomaly of the form \ref{eqn:anomform}, transforms under a 
conformal transformation. We can use this result to relate $W[B,{\bf h}]$ on
the perturbed FLRW space to $W[r,{\bf h}]$ on the perturbed four sphere
with radius $r$.
Writing $B(\chi ) = r\ e^{\sigma (\chi )}$, 
where $r$ is an arbitrary radius, the transformation is given by
\ba \label{acto4}
W [\sigma (\chi),h] & = & \tilde W [r,h] -
\frac{N^2}{32\pi^2} \int d^4 x \sqrt{\gamma} \left[ \sigma 
(R^{\mu \nu}R_{\mu \nu} - \frac{1}{3}R^2) 
+2 \nabla_{\mu} \sigma \nabla^{\mu} \sigma \nabla^2 \sigma \right.\nonumber\\
& & \qquad \qquad \qquad \left.
+2(R^{\mu \nu} - \frac{1}{2}\gamma^{\mu \nu} R)\nabla_{\mu} \sigma 
\nabla_{\nu} \sigma 
+(\nabla_{\mu} \sigma \nabla^{\mu} \sigma )^2 \right] 
\ea
Here $\tilde W$ denotes the effective action on the perturbed four sphere of 
radius $r$ with metric $\gamma_{\mu \nu}$, and the Ricci scalar $R$
and covariant derivative $\nabla_{\mu}$ refer to the same space.

The generating functional $\tilde W [r,h]$ was computed in 
\cite{Hawking00b,Hawking01a}, by using the AdS/CFT correspondence 
\cite{Maldacena98},
\be
\label{eqn:corres}
Z[{\bf h}] \equiv \int d[{\bf g}] \exp (-S_{grav}[{\bf g}]) = 
\int d[\phi] \exp (-S_{CFT}[\phi;{\bf h}]) \equiv 
\exp(-W_{CFT}[{\bf h}]),
\ee
where $Z[{\bf h}]$ is the supergravity partition function on $AdS_5$. 
The AdS/CFT calculation is performed by introducing a fictional ball of
(Euclidean) AdS that has the perturbed sphere as its boundary. In 
the classical gravity limit, the CFT generating functional can then be 
obtained by solving
the IIB supergravity field equations, to find the bulk metric ${\bf g}$ that 
matches onto the boundary metric ${\bf h}$, and adding a number of
counterterms that depend on the geometry of
the boundary, in order to render the action finite as the boundary is moved 
off to infinity. To second order in the perturbation ${\bf h}$,
the quantum effective action (including the $R^2$ counterterm) is given by
\be \label{Wper}
 \tilde W = \tilde W^{(0)} + \tilde W^{(1)} + \tilde W^{(2)} + \ldots
\ee
where 
\be\label{eqn:CFTgenfun0}
\tilde W^{(0)} = -\frac{3 \beta N^2 \Omega_4}{8 \pi^2} + \frac{3\alpha N^2
 \Omega_4}{4\pi^2} + \frac{3N^2 \Omega_4}{32 \pi^2} \left( 4\log 2 -1 \right), 
\ee
\be
\label{eqn:CFTgenfun1}
 \tilde W^{(1)} = \frac{3N^2}{16 \pi^2 r^2} \int d^4 x \sqrt{\hat{\gamma}} \, \psi,
\ee
\ba
\label{eqn:CFTgenfun2}
 \tilde W^{(2)} &=& -\frac{3 N^2}{64 \pi^2 r^4} \int d^4 x \sqrt{\hat{\gamma}}
\left[ \psi \left(\hat{\nabla}^2 +2\right) \psi - \alpha \psi
\left(\hat{\nabla}^4 + 4 \hat{\nabla}^2 \right) \psi \right] 
 \nonumber \\
 &+&  \frac{N^2}{256 \pi^2 r^4} \sum_p \left(\int d^4x'
 \sqrt{\hat{\gamma}} \, \theta^{\mu \nu}(x') H^{(p)}_{\mu\nu}(x') \right)^2 
\left( \Psi(p)  -4\alpha p(p+3)  \right) \nonumber,
\ea
where $p$ labels the eigenvalues of the Laplacian $\hat \nabla^2$
on the round four sphere and
\ba\label{cftten}
\Psi(p) & = &  p(p+1)(p+2)(p+3) \left[\psi(p/2+5/2) + \psi(p/2+2) -
\psi(2) - \psi (1)\right]\nonumber\\
& &  + p^4+2p^3-5p^2-10p -6 + 2\beta p(p+1)(p+2)(p+3),
\ea 
and we have allowed for a finite contribution, with coefficient $\beta$, 
of the third counterterm, which is necessary to cancel a logarithmic 
divergence of the tensor perturbation. Inserting the expression for
$\tilde W$ 
in \ref{acto4} and evaluating the terms depending 
on the scale factor $\sigma (\chi)$, one obtains 
the quantum effective action of the Yang-Mills teory on a 
general, perturbed FLRW geometry. For completeness, we also
give the Einstein-Hilbert action of the perturbed four sphere,
\ba\label{EHper}
S_{EH}& =& -\frac{3 \Omega_4 r^2}{4 \pi G} - \frac{3}{4 \pi G}
\int d^4 x \sqrt{\hat{\gamma}} \psi \nonumber\\
& &  + \frac{1}{16 \pi G r^2} \int d^4 x
\sqrt{\hat{\gamma}} \left( \frac{3}{2} \psi \hat{\nabla}^2 \psi 
+ 2 \theta^{\mu\nu} 
\theta_{\mu\nu} - \frac{1}{4} \theta^{\mu\nu} \hat{\nabla}^2
\theta_{\mu\nu} \right).
\ea

We shall use these results in section III, where we discuss the
instability of anomaly-induced inflation. 
But first, we return to the background evolution. In the next paragraph, 
we discuss a class of $O(4)$ invariant `real tunneling' instanton
solutions of the Starobinsky model \ref{tot} and study their role in
the no boundary path integral for the wave function of an inflationary
universe.

\subsection{Real Tunneling Geometries}

It is easily seen that the total action is stationary under
all perturbations $h_{\mu \nu}$, if the background is a round four sphere with 
radius
\be\label{dssol}
r^{2}_{s} = \frac{N^2 G}{4 \pi}.
\ee
By slicing the four sphere 
at the equator $\chi=\pi /2$ and writing $\chi = \frac{\pi}{2}-it$, it
analytically continues into the Lorentzian to the de Sitter 
solution mentioned above, with the cosmological constant provided by the trace
anomaly of the large $N$ Yang-Mills theory. 

Other compact, real instanton solutions of the form
\be\label{met}
ds^2 = d\tau^2 +b^2(\tau ) d\Omega_3^2
\ee
were found in \cite{Hawking01a}, by numerically integrating the
Einstein equations, which can be obtained directly 
from the trace anomaly by using energy-momentum conservation. 
Imposing regularity at the North Pole (at $\tau =0$) of the instanton
leaves only the third derivative of the 
scale factor at the North Pole as an adjustable parameter. 
It is convenient to define dimensionless variables $\tilde \tau =
\tau/r_{s}$ 
and $f(\tilde \tau) = b(\tau)/r_{s}$. For $\alpha < 0$, there exists 
a second regular, compact `double bubble' instanton, with $f'''(0)=-2.05$, 
together with a one-parameter family of instantons with an irregular South 
Pole. For $f'''(0)<-1$, the scale factor of the latter has two peaks.
For $-1 < f'''(0) < 0 $ on the other hand, they are similar to the
singular Hawking-Turok instantons that have been considered in
the context of scalar field inflation \cite{Hawking98}.

The Lorentzian part of the real tunneling saddle-points is obtained by
analytically continuing the instanton metric across a hypersurface of 
vanishing extrinsic curvature. The double bubble instanton can 
be continued across its `equator' to give a closed FLRW universe, or into an
open universe by a double continuation across the South Pole. Our 
numerical studies show that the closed universe rapidly collapses and that
the open spacetime hyper-inflates, with the scale factor blowing up at a 
finite time. Similarly, the singular instantons can be continued into an open 
FLRW universe across $\tau =0$, by setting $\tau=it$ and $\Omega_3 =i\phi$. 
For $f'''(0)<-1$ this again gives hyper-inflation, but for $-1 <f'''(0) < 0 $
one obtains a realistic inflationary universe. The four sphere solution
as well as the singular instantons that are small perturbations of $S^4$ at 
the regular pole, are most interesting for cosmology, since they yield long 
periods of inflation. 

Using the expressions \ref{acto4} and \ref{eqn:CFTgenfun0} for 
$W[\sigma (\chi)]$ and the relations
\be
\chi (\tau)=2\lim_{\epsilon \rightarrow 0}
\tan^{-1}\left[\tan(\epsilon/2)\exp\left(\int_{\epsilon}^{\tau} 
\frac{d\tau'}{b(\tau')}\right)\right], 
\quad \qquad
B(\tau)= \frac{b(\tau)}{\sin(\chi)},
\ee 
one can numerically compute the action of the real
tunneling geometries \cite{Hertog02}. 
On an unperturbed FLRW background, conformal to the round 
four sphere, the total Euclidean action becomes
\ba
S^{(0)}
& = & \frac{3 N^2 \Omega_3}{32 \pi^2} \int d\chi \sin^3 \chi \left[
\frac{1}{3}(12 (\log 2+\sigma -\beta)-3 +6 \sigma'^2 - \sigma'^4
-4\sigma'^3 \cot \chi) \right. \nonumber\\
& & \left. \qquad -e^{2 \sigma}(\sigma'^2 +2) 
+ 2\alpha (\sigma''+3 \sigma' \cot \chi +\sigma'^2 -2)^2\right]
\ea
where $\sigma = \log (B/r)$.
On the round four sphere, $\sigma \rightarrow 0$, so the action reduces to
\be \label{acto5}
S^{(0)} =\frac{3 N^2 \Omega_4}{32 \pi^2}(8\alpha -3 +4 (\log 2 -\beta))
\ee 

We find that for all $\alpha <0$ the regular double bubble instanton has 
much lower action than the four sphere. The singular double bubble 
instantons have divergent action, but the Hawking-Turok type instantons have 
finite action\footnote{For completeness, we should mention that if 
$\alpha > 0$ one must have $f'''(0)\leq -1$ in order 
for the solution to be compact. 
For $f'''(0) < -1$, the instantons have a singular South Pole but finite 
action, and continue to hyper-inflating open universes.}. For given $\alpha$, 
the action of the latter class 
depends on the third derivative of the scale factor at $\tau=0$. This is the
analogue of the situation in scalar field inflation, 
where the action of the Hawking-Turok instantons
depends on the value of the inflaton field at the North Pole. 
The action of the singular instantons tends smoothly to  
the $S^4$ action \ref{acto5} as $f''' (0) \rightarrow -1$ and 
it decreases monotonically with increasing $f''' (0)$. 

To summarize, we found a one-parameter family of finite-action, 
compact solutions of the Euclidean field equations that can be
analytically continued across a spacelike surface 
$\Sigma$ of vanishing curvature, to Lorentzian geometries that describe  
realistic inflationary universes. The condition on
$\Sigma$ guarantees that a real solution of the Euclidean field equations is 
continued to a real Lorentzian spacetime. The Euclidean region is essential, 
since there is no way to round off a Lorentzian geometry without introducing 
a boundary. What is the relevance then, in the context of the no boundary
proposal, of these real tunneling geometries with regard to the problem 
of initial conditions in cosmology?

At least at the semiclassical level, the no boundary proposal gives a measure 
on the space of initial conditions for cosmology. The weight of each classical
trajectory is approximately $\vert \Psi \vert^2 \sim
e^{-2S_R}$, where $S_R$ is the real part of the Euclidean action
of the solution. For real tunneling solutions this comes entirely
from the part of the manifold on which the geometry is Riemannian.
The simplicity of this situation  
has led to the interpretation of the no boundary proposal 
as a bottom up theory of initial conditions.
In particular, it has been argued that if a given theory allows different
instantons, the no boundary proposal predicts our universe to be created 
through the lowest-action solution, since this would give the dominant
contribution to the path integral. Applying this interpretation to trace 
anomaly driven inflation, one must conclude that the no boundary proposal 
predicts the creation of a hyper-inflating universe emerging from the double 
bubble instanton, or a nearly empty open universe that occurs by semiclassical
tunneling via a singular instanton with $\vert f'''(0) \vert $ small. 

The situation is similar in many theories of scalar field inflation. 
Restricting attention to real tunneling geometries, a
bottom up interpretation of the no boundary proposal generally favours the 
creation of large spacetimes. One typically obtains a probability distribution 
that is peaked around instantons in which the field at the surface of 
continuation is near the minimum of its potential, yielding very little 
inflation. Hence, the most probable universes are nearly empty open universes 
or collapsing closed universes, depending on the analytic continuation one 
considers. Weak anthropic arguments have been invoked to try to rescue the 
situation \cite{Hawking98}, by weighing the a priori no boundary probability 
with the probability of the formation of galaxies. However, for the most 
natural inflaton potentials, this still predicts a value of $\Omega_0$ that is 
far too low to be compatible with observations. Another attempt \cite{Turok00},
based on introducing a volume factor that represents the projection onto the 
subset of states containing a particular observer, leads to eternal inflation 
at the Planck density, where the theory breaks down. In fact, invoking 
conditional probabilities is contrary to the whole idea of the no boundary 
proposal, which by itself specifies the quantum state of the universe.

Clearly the predictions of a bottom up interpretation of the no boundary
proposal do not agree with observation.
This is because it is an essentially classical interpretation,
which is neither relevant nor correct for cosmology.
The quantum origin of the universe implies its quantum state
is given by a path 
integral. Therefore, one must adopt a quantum approach to the problem
of initial conditions, in which one considers the
no boundary path integral \ref{nbp} for a given quantum state of the 
universe. We shall apply such a quantum approach 
in section IV, to describe the origin of an inflationary universe, in
theories like trace anomaly inflation.
It turns out that the relevant saddle-points are not exactly 
real tunneling geometries. Instead, one must consider
complex saddle points, in which the geometry becomes gradually 
Lorentzian at late times.

\section{Instability of Anomaly-Induced Inflation}

\subsection{Metric Perturbations}

Two-point functions of metric perturbations can be computed directly from 
the no boundary path integral. One perturbatively evaluates the path
integral around an $O(4)$ invariant instanton background to obtain
the real-space Euclidean 
correlator, which is then analytically continued into the Lorentzian 
universe, where 
it describes the quantum fluctuations of the graviton field in the 
primordial de Sitter phase \cite{Gratton99,Hertog00}.
The quantum state of the Lorentzian 
fluctuations is uniquely determined by the condition of regularity 
on the instanton \cite{Hawking01a}. 
Both scalar and tensor perturbations are given by a path integral of
the form
\be \label{path}
\langle h_{\mu\nu}(x) h_{\mu'\nu'}(x') \rangle \sim
\int d[{\bf h}] \exp (-S^{(2)})h_{\mu \nu}(x) h_{\mu'\nu'}(x'), 
\ee 
where $S^{(2)}$ denotes the second order perturbation of the action 
\be \label{actper}
S  = S_{EH} +S_{GH} +S_{R^2} +\tilde W,
\ee
with $\tilde W$ given by \ref{Wper}.
For the scalars, eliminating the remaining gauge freedom 
introduces Faddeev-Popov ghosts. These ghosts supply 
a determinant $(\hat \nabla^2 +4)^{-1}$, which cancels a similar factor
in the scalar action, rendering
it second order\footnote{The gauge freedom also leads to closed loops 
of Faddeev-Popov ghosts but they can be neglected in the 
large $N$ approximation.}. The action for the tensors $\theta_{\mu\nu}$ on 
the other hand is non-local and fourth order.
Nevertheless, the metric perturbation and its first derivative should not
be regarded as two independent variables, since 
this would lead to meaningless probability distributions 
in the Lorentzian \cite{Hawking01b}.
Instead the path integral should be taken over the fields $\theta_{\mu\nu}$
only\footnote{This means one loses unitarity.
However, probabilities for observations tend towards those 
of the second order theory, as the coefficients of the fourth order terms in 
the action tend to zero. Hence unitarity is restored at the low energies that 
now occur in the universe.}, to compute correlators of the form \ref{path}. 
The Euclidean action for $\theta_{\mu\nu}$ is positive definite, so the path 
integral over all $\theta_{\mu\nu}$ converges and determines a well-defined 
Euclidean quantum field theory. One might worry that the higher derivatives 
would lead to instabilities in the Lorentzian. This is not the case, however, 
since the no boundary prescription to compute 
Lorentzian propagators by Wick rotation from the Euclidean, implicitly
imposes the final boundary condition that the fields remain bounded, 
which eliminates the runaways \cite{Hawking01a,Hawking01b}.

The path integral \ref{path} is Gaussian, so the correlation functions 
can be read off from the perturbed action, equation
\ref{eqn:CFTgenfun2} and \ref{EHper}:
\be
\label{eqn:scalarcorrelator}
\langle \psi (x) \psi (x') \rangle = \frac{32 \pi^2 r^4_{s}}{3
\vert \alpha \vert N^2} \left( -\hat{\nabla}^2 + 1/2\alpha \right)^{-1},
\ee
and 
\be
\label{eqn:tensorcorrelator}
\langle \theta_{\mu\nu}(x) \theta_{\mu'\nu'}(x') \rangle = 
\frac{128 \pi^2  r^4_{s}}{N^2} \sum_{p=2}^{\infty}
\frac{W^{(p)}_{\mu\nu\mu'\nu'}(x,x')}{p^2 + 3p +6 + \Psi(p)- 4\alpha p(p+3)},
\ee
where the bitensor $W^{(p)}_{\mu\nu\mu'\nu'}(x,x')$ is defined as the usual 
sum over degenerate rank-2 harmonics on the four sphere and $\Psi (p)$ is 
given by \ref{cftten}.

The scalar two-point function \ref{eqn:scalarcorrelator} is just the 
propagator of a particle with physical mass $m^2 =(2 \alpha r^2_{s})^{-1}$. 
Since we are assuming $\alpha < 0$, we have $m^2 < 0$ so this particle is a 
tachyon, which is the perturbative manifestation 
of the Starobinsky instability.
Making $\alpha$ more negative, makes the tachyon mass squared less negative,
and therefore weakens the instability. Indeed, the number of efoldings
in the primordial de Sitter phase emerging from the four sphere instanton
is given by $N_{efolds} \sim 12 \vert \alpha \vert (\log N -1)$. Therefore,
in the interesting regime, we have $-m^2 << m^2_{pl}$, so semiclassical
gravity should be a good approximation. 

This result sheds light on the problem of initial conditions in trace anomaly 
inflation. One can think of the non derivative term in the scalar 
correlator as a potential $V(\psi)$, with the unperturbed de Sitter 
solution at $\psi =0$ at the maximum. 
If $\vert \alpha \vert$ is not too small, 
then the top of the potential is sufficiently flat, so that the lowest-action 
regular instanton is a homogeneous Hawking-Moss instanton
\cite{Hawking82a}, with $\psi$ constant at the top.
Since the instability of the de Sitter phase is
characterised entirely by the coefficient $\alpha$ of the $R^2$ counterterm.
this means the problem of initial 
conditions in anomaly-induced inflation is similar to the 
corresponding problem in many theories of scalar field
inflation, where one ought to explain why the inflaton starts
initially at the top of the hill. 
We study the origin of these inflationary universes in section IV.
Before doing so, however, we comment on the homogeneous mode in the scalar
spectrum, which has given rise to some controversy in the literature.
 
\subsection{Homogeneous Fluctuations}

The most interesting instantons in both trace anomaly driven inflation 
as well as most theories of scalar field inflation possess a homogeneous 
fluctuation mode which decreases their action \cite{Hawking01a,Gratton01}. 
The presence of such a negative mode is the perturbative manifestation of the 
conformal factor problem. Indeed,
since the conformal factor problem is closely related to the instability of
gravity under gravitational collapse, one expects instantons that 
are appealing from a cosmological perspective, to possess a negative mode. 

Writing the scalar propagator \ref{eqn:scalarcorrelator}
on the four sphere instanton in momentum space gives
\be \label{eqn:scor}
 \langle \psi(x) \psi(x') \rangle = \frac{32 \pi^2 r^4_{s}}{3|\alpha | N^2}
\sum_{p=0}^{\infty} \frac{W^{(p)}(\mu(x,x'))}{p(p+3)+m^2},
\ee
where the biscalar $W^{(p)}$ equals the usual 
sum over degenerate scalar harmonics on the four sphere with eigenvalue
$\lambda_{p} = -p(p+3)$ of the Laplacian.

There are many negative modes if $ -1/8< \alpha <0$. This is usually the 
perturbative indication of the existence of a lower-action instanton solution.
For instance, in scalar field inflation with a double well potential,
the Hawking-Moss instanton possesses several negative modes if 
$V_{,\phi \phi}/H^2 <-4$, which is precisely the condition for the existence 
of a lower-action Coleman-De Luccia instanton that straddles the maximum.
On the other hand, if $\alpha < -1/8$ in \ref{eqn:scor} then only the 
homogeneous ($p=0$) negative mode remains, which is again similar to the 
well-known negative homogeneous mode of the Hawking-Moss instanton 
in theories with a scalar potential that is sufficiently flat. 

The presence of a physical negative mode supports the interpretation of an 
instanton as describing the decay of an unstable state through semiclassical 
tunneling \cite{Coleman80}. On the other hand, it has been argued 
that it questions its use in the no boundary path integral to define the 
initial quantum state of the universe\footnote{In scalar field
inflation, one can view the singular Hawking-Turok instantons as
constrained instantons, with additional data specified on an internal 
boundary. For some theories, the constraint introduced in
\cite{Kirklin01} to resolve the singularity, also removes the 
negative mode, at least perturbatively \cite{Gratton01}.
However, it does not remove the instability non-perturbatively and 
for the most obvious potentials, the lowest-action constrained instanton
gives very little inflation.} 
\cite{Gratton01}. Within the semiclassical approximation, however,
it is more appropriate to project out the negative mode,
since the semiclassical approach is based on the {\em assumption} that the 
path integral can be expanded around solutions of the classical field 
equations. 

The conclusions of \cite{Gratton01} are based on a 
perturbation calculation around compact, real instanton backgrounds, 
that does not take in account the wave function of interest.
One expects, however, the configuration specifying the
quantum state of the Lorentzian universe 
to project out the negative mode from the perturbation spectrum.
Consider for example the wave function of a universe described by a
3-sphere with radius $R^2 = V_{0}/3$ and field $\phi=0$, in a theory of
gravity coupled to a single scalar field with potential $V_0 (1-\phi^2)^2$.
In the semiclassical approximation, this is given by half of a
Hawking-Moss instanton with the field constant at the top of the potential.
Obviously, this solution has no negative mode, since the 
boundary condition on the 3-sphere 
$\Sigma$ removes the lowest eigenvalue 
solution of the Schr\"{o}dinger equation for the perturbations. 
Since the negative mode corresponds to 
a homogeneous fluctuation, this is probably
true also for large 3-spheres in the Lorentzian regime.
Therefore, one expects that in the top down approach to cosmology, where
the quantum state of the universe is taken in account, the negative
mode is automatically projected out.

\subsection{Quantum Matter and the Microwave Background}

Before discussing the top down approach in more detail, we pause to briefly 
comment on some of the characteristic predictions for observations
of trace anomaly inflation.

To extract accurate predictions for the cosmic microwave anisotropies,
one must evolve the perturbations through the Starobinsky 
instability, to obtain initial conditions for the inhomogeneities during the
radiation and matter eras. Details of this calculation will be
presented elsewhere \cite{Hawking02}, but some interesting
features of the microwave temperature anisotropies predicted
by anomaly-induced inflation, can be extracted from the correlators 
\ref{eqn:scalarcorrelator} and \ref{eqn:tensorcorrelator} in the
primordial de Sitter era. Obviously, as can be seen from 
\ref{eqn:tensorcorrelator}, the quantum matter couples to the 
tensors. Starobinsky \cite{Starobinsky83} and Vilenkin \cite{Vilenkin85} 
assumed that the amplitude of primordial gravity waves was not 
significantly altered by the quantum matter loops.
This assumption can now be examined using AdS/CFT, which has allowed us to
include the effect of the Weyl$^2$ counterterm and the 
non-local part of the matter effective action.
We find that at small scales, matter fields dominate the tensor 
propagator and make it decay like $p^4 \log p$.
In other words, the CFT appears to give spacetime a
rigidity on small scales, an example of how quantum loops of matter
can change gravity at short distances. In fact, this suppression
should occur even if inflation is not driven by the trace anomaly,
since we observe a large number of matter fields, whose effective
action is expected to dominate the propagator on small scales.

Secondly, both the higher derivative counterterms and the matter fields 
introduce anisotropic stress, which is an important difference with 
scalar field inflation. This can be seen from decomposing the tensors
$\theta_{\mu\nu}$ into a scalar $\phi$ and tensor $t_{ij}$
under $O(4)$. The former is the difference between the two potentials 
in the Newtonian gauge and corresponds to anisotropic stress.
Typically reheating at the end of
anomaly-induced inflation leads to creation of particles 
that are not in thermal equilibrium with the photon-baryon fluid, so
one expects some anisotropic stress to survive during the radiation era.
To make more precise predictions, however, a better understanding is 
required of the (probably time-dependent) values of the coefficients
$\alpha$ and $\beta$ of the higher derivative counterterms in the theory.

Finally, we should mention that for the tensor propagator the
higher derivative terms also give rise to poles in the complex $p$-plane.   
These are harmless, however, since the contour obtained from the
Euclidean goes around the complex poles \cite{Hawking01a}.
In other words, defining our theory in the Euclidean, implicitly removes the
instabilities associated with the complex poles, 
like a final boundary condition removes the runaway solution of the
classical radiation reaction force \cite{Hawking01b}. 

\section{Origin of Inflation}

We have seen that the predictions of the bottom up approach to the problem 
of initial conditions in inflation do not agree with observation.
This is because it is based on an essentially classical picture,
in which one assumes some initial condition
for the universe and evolves it forward in time.
The quantum origin of our universe, however, means that its wave 
function is determined by a path integral,
in which one sums over all possible histories that lead to a given 
quantum state, together with some suitable boundary conditions 
on the paths. This naturally leads to a top down view of the 
universe. In a top down context, rather than comparing the relative 
probabilities of different semiclassical geometries, one looks for 
the most probable evolution that leads to a certain outcome.

We now apply the quantum top down interpretation of the no boundary proposal
to study the origin of an inflationary universe, in theories where 
the instability of the inflationary phase can be described in terms of 
a single scalar field with an effective potential that has a local maximum.
As shown in section III, this includes trace anomaly driven inflation,
since the emergence of an anomaly driven inflationary universe is very similar 
to the creation of an exponentially expanding universe in theories of 
new inflation. 

We consider a model consisting of gravity coupled to a single scalar field, with 
a double well potential $V(\phi) = A(1-\frac{C}{2}\phi^2)^2$ (with $A,C >0$).
For $C<2/3$, the potential has a maximum at $\phi=0$
with $V_{,\phi \phi}/V$ sufficiently low so that there exists no 
Coleman-De Luccia instanton, but only a 
Hawking-Moss instanton with $\phi=0$ everywhere on top of the hill.
Implementing a top down approach, we consider the quantum amplitudes 
$\Phi [\Sigma,\tilde {\bf h},K,\phi_{\Sigma}]$ for 
different conformal 3-geometries $\tilde {\bf h}$ with trace
$K$ of the second fundamental form, on an
expanding surface $\Sigma$ during inflation\footnote{In principle we should 
consider the amplitude for a conformal 3-geometry on a surface $\Sigma$ just 
inside our past light cone, with $K$ equal to
the present Hubble rate and given values for all other observables. 
However, it is sufficient to consider the quantum amplitude for a 
configuration on an expanding surface in the inflationary period,
since this can then be accurately evolved to the future using classical laws.}.
According to the no boundary proposal, the defining path integral should be
taken over all compact Riemannian geometries that induce the prescribed
configuration on $\Sigma$. 

In the $K$-representation, the Euclidean action is given by
\be\label{totnew}
 S = -\frac{1}{2\kappa} \int d^4 x \sqrt{g} R - \frac{1}{3\kappa}
\int d^3 x \sqrt{h} K + \int d^4 x \sqrt{g} \left( \frac{1}{2}g^{\mu\nu}
\partial_{\mu} \phi \partial_{\nu} \phi + V(\phi)\right),
\ee
The usual wave function $\Psi [{\bf h},\phi_{\Sigma}^{\ }]$ is
obtained from $\Phi [\tilde {\bf h},K,\phi_{\Sigma}]$ by an
inverse Laplace transform,
\be
\Psi [{\bf h},\phi_{\Sigma}^{\ }]=
\int_{\Gamma} d\left[\frac{K}{4i\kappa}\right] 
\exp \left[\frac{2}{3\kappa} \int d^3 x 
\sqrt{h} K \right] \Phi [\tilde {\bf h},K,\phi_{\Sigma}]
\ee
where the contour $\Gamma$ runs from $-i\infty$ to $+i\infty$.

Within the semiclassical approximation, the no boundary wave function
is approximately given by the saddle-point contributions.
Restricting attention to saddle-points that are invariant under the action
of an $O(4)$ isometry group, the instanton metric can be written as
\be\label{met2}
ds^2 = d\tau^2 +b^2(\tau ) d\Omega_3^2,
\ee
and the Euclidean field equations read
\be \label{efield1}
\phi'' =  -K\phi' +V_{,\phi}
\ee
\be\label{efield2}
K' +K^2=  -(\phi^2_{,\tau} +V )
\ee
where $\phi'=\phi_{,\tau}$ and $K= 3b_{,\tau}/b$.
The Lorentzian trace $K_{L}^{\ }=-3\dot a/a $ is obtained by analytic 
continuation. We first calculate the wave function for real $K$, 
and then analytically continue to imaginary, or Lorentzian 
$K_{L}^{\ }= -iK$.
 
At the semiclassical level, there are two contributions to the given amplitude.
For small $\phi_{\Sigma}$ and any Euclidean $K$, there always exists a 
non-singular, Euclidean $O(4)$ invariant solution of the field equations, with 
the prescribed boundary conditions. This solution is part of a deformed 
sphere, or Hawking-Turok instanton. In the approximation $K=3H\cot (H\tau)$, 
with $H^2 = A/3$, and $V(\phi) \sim A(1-C\phi^2)$,
the solution of \ref{efield1} is given by
\be
\phi = \phi_{\Sigma}^{\ } \frac{\ _2 F_1(3/2 +q,3/2-q,2,z( K))} 
{\ _2 F_1(3/2 +q,3/2-q,2,z( K_{\Sigma}))}
\ee
where 
\be
q = \sqrt{9/4 +6C},  
\qquad \quad
z( K) =  \frac{1}{2}
\left[1-\frac{ K}{(A^2+ K^2)^{1/2}}\right]
\ee
At the South Pole $K \rightarrow +\infty$, so in the instanton the scalar 
field slowly rolls up the hill from its value at the regular South Pole to the 
prescribed value $\phi_{\Sigma}$ on the 3-sphere with trace $K_{\Sigma}^{\ }$. 
The weight of the Hawking-Turok geometry in the no boundary path
integral for the wave function $\Phi [K,\phi_{\Sigma}^{\ }]$ 
is approximately given by
\ba \label{pert}
S[K_{\Sigma}^{\ }, \phi_{\Sigma}^{\ }] & = & - \frac{1}{3\kappa}
\int d^3 x \sqrt{h} K -\int d^4 x \sqrt{g}  V(\phi) \nonumber\\
&  = & - \frac{12\pi^2}{A}
\left[1-\frac{ K}{(A^2+ K^2)^{1/2}}\right]
-\frac{24\pi^2 C}{A^2(1-C)}\phi_{\Sigma}^2 \  z^2(K_{\Sigma}^{\ })
\times \nonumber\\
& & \left[1-2 z(K_{\Sigma}^{\ })
+3C(1-z(K_{\Sigma}^{\ }))\left(z(K_{\Sigma}^{\
})+\frac{2-3C}{1+3C}\right) \frac{F'}{F}[(z(K_{\Sigma}^{\ })] \right]
\ea

For small $\phi_{\Sigma}^{\ }$, there is a second semiclassical
contribution to the wave function, coming 
from universes that are created via an $O(5)$ symmetric Hawking-Moss 
instanton with $\phi$ constant at the top of the hill, but in which
a quantum fluctuation disturbs the field, 
causing it to run down to its prescribed value $\phi_{\Sigma}^{\ }$ at 
the 3-sphere boundary with trace $K_{\Sigma}^{\ }$.
Neglecting prefactors, the action of the Hawking-Moss geometry
is given by the first term in \ref{pert}. 
It follows that for $K_{\Sigma}^{\ }=0$, the action of
the Hawking-Turok geometry is 
more negative than the action of the Hawking-Moss
instanton. This would seem to suggest that the universe is least 
likely to start at the top of the hill. However, we are not interested in the 
amplitude for a Euclidean spacetime, but in the no boundary wave
function of a Lorentzian expanding universe. 

Within the regime in which $\phi$ remains small over the whole geometry, 
one can derive the amplitude in the Lorentzian from our result for 
$S[K_{\Sigma}^{\ }, \phi_{\Sigma}^{\ }]$, by analytic continuation
into the complex $K$-plane. In a Lorentzian universe, Euclidean $K$ is pure 
imaginary, $K_{L}^{\ }= -iK$. Since the
action is invariant under diffeomorphically related contours in the complex 
$\tau$-plane, we may deform the contour into one with straight
sections, along the real and imaginary $K$-axis. 
It follows immediately from \ref{pert} that the real part of the 
action for the Hawking-Moss instanton is constant on the imaginary $K$-axis,
unlike the action for the Hawking-Turok geometry. 
According to the no boundary proposal, the relative probability 
of both geometries is given by
\be
P [K_{L},\phi_{\Sigma}] = \frac{A_{HM}^2}{A_{HT}^2}
e^{-2\Re[\Delta S]}
\ee
where $\Delta S = S_{HM}^{\ }-S_{HT}^{\ }$.
The prefactors account for small fluctuations around the classical
solutions and can be neglected for small $\phi$.

\begin{figure}
\centerline{\psfig{file=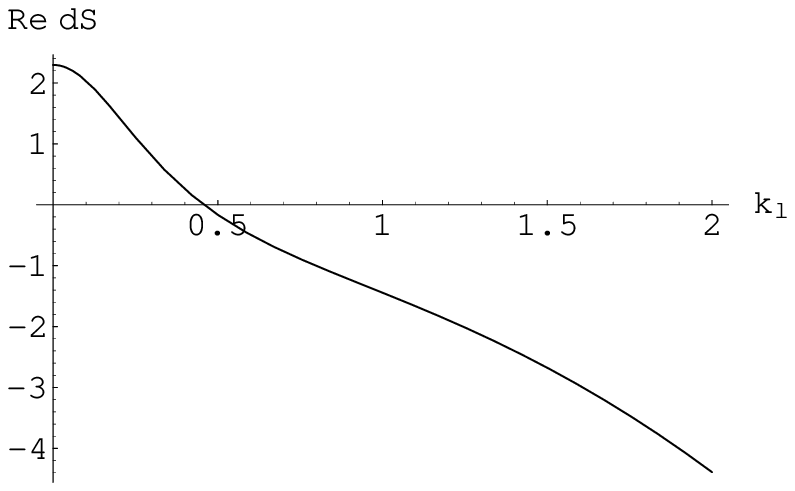,width=4.in}}
\caption{$\Re[\delta S]$ is proportional to the difference $\Delta S$ 
between the action of the Hawking-Moss and Hawking-Turok type geometries
discussed in the text. 
\newline
The no boundary proposal predicts an expanding universe 
to be created in an unstable de Sitter state, by
semiclassical tunneling via a Hawking-Moss instanton.}
\end{figure}

In figure 1 we plot $\Re[\delta S (k_l)]$, which is proportional to
the real part of the
difference $\Delta S$ between the action of both geometries, as a function of 
Lorentzian $k_{l} = \frac{K_{L}}{\sqrt{A^2-K^2_{L}}}$, with $A=1$ and $C=1/3$.
This shows that the real part of the action for 
the Hawking-Turok geometry increases on the imaginary axis away from $K=0$
and soon becomes larger than the $O(5)$ action. 
In addition, within our approximation $\phi_{\Sigma}$ enters only in
the prefactor of $\Delta S$.
Therefore, the dominant contribution to the no boundary path integral for a
Lorentzian inflating universe comes from spacetimes which are
created by semiclassical tunneling via a Hawking--Moss instanton
and which inflate for a long time before a quantum fluctuation 
causes the field to roll down to its final value $\phi_{\Sigma}$.
This means that in an inflationary universe, the scalar field is more likely 
to start at the top of the hill and roll down, than to start lower down. 
The reason is that although being at the top of the hill costs potential 
energy, it saves gradient energy, by having a scalar field that is constant in
space and time. If the maximum of the potential is fairly flat, the gradient
energy is dominant, and the universe starts with a constant scalar, at
the top of the hill. Therefore, one does not need an initial hot Big 
Bang phase, to explain why inflation began at a local maximum of the potential.

As mentioned above, this scenario is realised in trace anomaly driven inflation.
The unperturbed de Sitter solution \ref{dssol} in anomaly-induced inflation
emerges from the Hawking-Moss geometry, while the inhomogeneous 
Hawking-Turok evolution
corresponds to one of the singular instantons discussed in section II.
The field configuration on $\Sigma$ determines the 
third derivative of the scale factor at the regular South Pole, or equivalently
the initial value of the order parameter $\phi$ governing the instability.
For $\alpha <-1/8$, the instability of the de Sitter phase is sufficiently weak, 
so that the universe is most likely to start at the top,
in an unstable de Sitter state. 
This result also justifies our calculation of metric perturbations, 
which were based on a perturbative expansion of the path integral
about the round four sphere.

Finally, we should mention that because we are interested in real 
matter fields on $\Sigma$, the analytic continuation into the complex
$K$-plane means $\phi$ must be complex in the bulk of the instanton\footnote{Complex 
instanton solutions have previously been considered  
in \cite{Lyons92}. Physical constraints on complex
contours of 4-geometries
in the no boundary path integral were discussed 
in \cite{Halliwell90a}. In this context, we should mention that in
the absence of an extension of Bishop's theorem to complex geometries, 
it is not entirely clear whether the $O(4)$ invariant geometries considered 
here are in fact the lowest-action saddle-points that contribute to
the wave function of interest.}. 
More precisely, at the South Pole, we must have
$\Im [\phi] = \phi_{\Sigma} \Im [F(z_{\Sigma})]/\Re[F(z_{\Sigma})]$.
This has no physical meaning though,
since the stationary phase approximation is just a mathematical construction
to evaluate the path integral over real $\phi$.

\section{Discussion}

We have studied the problem of initial conditions in cosmology.
Because our universe has a quantum origin, one must adopt
a top down approach, in which one considers the path
integral over a class of histories
that lead to a given quantum state of the universe.
A top down view is naturally implemented in the context of the no boundary
proposal, which states that the amplitude for the quantum state
of the universe on a 3-surface $\Sigma$ is given by a path integral
over all geometries that induce the prescribed configuration on their only
boundary $\Sigma$. We have investigated the no boundary predictions for the 
quantum cosmological origin of a large class of inflationary universes.
In particular, we have considered theories of inflation where the global
instability can be described in terms of a single scalar field $\phi$ with an
effective potential that has a sufficiently flat local maximum.
This includes Starobinsky's trace anomaly model, since the nature of the 
initial instability in anomaly-induced inflation is the same as the
instability that occurs in new inflation. Trace anomaly driven inflation has
a sound motivation in fundamental particle physics, since we observe a 
large number of matter fields in the universe, which may be expected to 
behave like a CFT in the early universe.
The no boundary proposal predicts an inflationary
universe to be created in an unstable de Sitter state, 
by semiclassical tunneling via a Hawking-Moss instanton. 
The universe first inflates, before a quantum fluctuation 
causes the field to roll down and inflation to end. 
Provided $-4 < V_{,\phi \phi}/H^2 <0$, the
maximum of the potential is sufficiently flat, so that this geometry has 
lower action than an inhomogeneous Hawking-Turok type evolution. 

One could think of the no boundary proposal as describing the creation of 
universes with different radii, like the formation of bubbles of steam. 
If the bubbles are small, they collapse again, but there is a 
critical size above which they are more likely to grow. 
In theories where the amplitude for an expanding universe is
dominated by geometries that start in a de Sitter state, this naturally 
leads to the interpretation of the round Hawking-Moss instanton on top
of the hill as corresponding to that critical size.

Correlators of observables on a spacelike
hypersurface $\Sigma$ should be computed directly from the no boundary
path integral, by summing over histories to the past of that surface.
In the semiclassical approximation, the 
dominant instanton saddle-point solution should be used as background in a 
perturbative evaluation of the no boundary path integral, to find correlators
of metric perturbations during inflation. Therefore, our result
justifies the perturbation calculations performed in section III and 
in \cite{Hawking01a}, in which we computed Euclidean propagators 
assuming a four sphere instanton
background. Evolving the spectrum of primordial perturbations through the
Starobinsky instability determines the cosmic microwave anisotropies. 
Hence the boundary condition on the fluctuation
modes imposed by the instanton background may provide an
observational discriminant between different saddle-points, hereby
connecting quantum cosmology and the top down approach 
to falsifiabble predictions for observation \cite{Gratton00}. 

We have argued that because our universe has a quantum origin, one must adopt 
a top down approach to the problem of initial conditions in cosmology,
in which the histories that contribute to 
the path integral, depend on the observable being measured. 
There is an amplitude for empty flat space, but it is not of much significance.
Similarly, the other bubbles in an eternally inflating spacetime are irrelevant. 
They are to the future of our past light cone, so they don't contribute to the 
action for observables and should be excised by Ockham's razor. 
Therefore, the top down approach is a mathematical formulation of the weak 
anthropic principle. Instead of starting with a universe and asking what a 
typical observer would see, one specifies the amplitude of interest.
In the context of the no boundary proposal, however, 
a top down description of the universe 
is not necessarily less `complete' or less predictive. 
We believe that if we are to explain why the universe is the way we 
observe it to be, a top down view is forced upon us by the quantum nature of 
the universe. Therefore, although future developments in M-Theory will provide 
us with new insights in how a theory of boundary conditions in cosmology 
should be formulated, the approach developed here should still apply when 
the framework of quantum cosmology will be based on M-Theory.

\bigskip

\centerline{{\bf Acknowledgments}}

It is a pleasure to thank Gary Gibbons, Harvey Reall and Toby Wiseman for 
helpful discussions.

\end{document}